\documentclass[iop]{emulateapj}

\usepackage{graphicx}
\usepackage{amsmath}
\usepackage{natbib}
\usepackage{mathptmx}
\usepackage{apjfonts}
\usepackage{times}
\usepackage[colorlinks=true,urlcolor=blue,citecolor=blue,linkcolor=blue]{hyperref}
\usepackage{aas_macros}

\bibpunct{(}{)}{;}{a}{}{,}

\usepackage[position=t,singlelinecheck=on,font={rm,bf,up},font=large]{subfig}

\makeatletter
\long\def\frontmatter@title@above{
  \vspace*{-17mm}\vspace*{\headheight}
   \hspace{-3mm}{\sc The Astrophysical Journal Supplement Series}, 229:9 (12pp), 2017\\
   \vspace*{4mm}{\footnotesize {\sc Preprint typeset using \LaTeX\ style emulateapj}}
  \par\vspace*{-\baselineskip}\vspace{6mm}
  }

\makeatother

\RequirePackage{color}

\shorttitle{Transverse oscillations in slender Ca~{\sc ii}~H fibrils}
\shortauthors{Jafarzadeh et al.}

\begin{document}

\title{Transverse Oscillations in Slender Ca~{\sc ii}~H Fibrils Observed with {\sc Sunrise}/SuFI}

\author{S.~Jafarzadeh\hyperlink{}{\altaffilmark{1}}}
\author{S.~K.~Solanki\hyperlink{}{\altaffilmark{2,3}}}
\author{R.~Gafeira\hyperlink{}{\altaffilmark{2}}}
\author{M.~van~Noort\hyperlink{}{\altaffilmark{2}}}
\author{P.~Barthol\hyperlink{}{\altaffilmark{2}}}
\author{J.~Blanco~Rodr\'{i}guez\hyperlink{}{\altaffilmark{4}}}
\author{ J.~C.~del~Toro~Iniesta\hyperlink{}{\altaffilmark{5}}}
\author{A.~Gandorfer\hyperlink{}{\altaffilmark{2}}}
\author{L.~Gizon\hyperlink{}{\altaffilmark{2,6}}}
\author{J.~Hirzberger\hyperlink{}{\altaffilmark{2}}}
\author{M.~Kn\"{o}lker\hyperlink{}{\altaffilmark{7,}\altaffilmark{9}}}
\author{D.~Orozco~Su\'{a}rez\hyperlink{}{\altaffilmark{5}}}
\author{T.~L.~Riethm\"{u}ller\hyperlink{}{\altaffilmark{2}}}
\author{W.~Schmidt\hyperlink{}{\altaffilmark{8}}}

\affil{\altaffilmark{1}\hspace{0.2em}Institute of Theoretical Astrophysics, University of Oslo, P.O. Box 1029 Blindern, N-0315 Oslo, Norway; \href{mailto:shahin.jafarzadeh@astro.uio.no}{shahin.jafarzadeh@astro.uio.no}\\
\altaffilmark{2}\hspace{0.2em}Max Planck Institute for Solar System Research, Justus-von-Liebig-Weg 3, 37077 G\"{o}ttingen, Germany\\
\altaffilmark{3}\hspace{0.2em}School of Space Research, Kyung Hee University, Yongin, Gyeonggi 446-701, Republic of Korea\\
\altaffilmark{4}\hspace{0.2em}Grupo de Astronom\'{\i}a y Ciencias del Espacio, Universidad de Valencia, 46980 Paterna, Valencia, Spain\\
\altaffilmark{5}\hspace{0.2em}Instituto de Astrof\'{i}sica de Andaluc\'{i}a (CSIC), Apartado de Correos 3004, E-18080 Granada, Spain\\
\altaffilmark{6}\hspace{0.2em}Institut f\"ur Astrophysik, Georg-August-Universit\"at G\"ottingen, Friedrich-Hund-Platz 1, 37077 G\"ottingen, Germany\\
\altaffilmark{7}\hspace{0.2em}High Altitude Observatory, National Center for Atmospheric Research, \footnote{The National Center for Atmospheric Research is sponsored by the National Science Foundation.} P.O. Box 3000, Boulder, CO 80307-3000, USA\\
\altaffilmark{8}\hspace{0.2em}Kiepenheuer-Institut f\"{u}r Sonnenphysik, Sch\"{o}neckstr. 6, D-79104 Freiburg, Germany}

\altaffiltext{9}{The National Center for Atmospheric Research is sponsored by the National Science Foundation.}

\begin{abstract}

We present observations of transverse oscillations in slender Ca~{\sc ii}~H fibrils (SCFs) in the lower solar chromosphere. We use a 1~hr long time series of  high- (spatial and temporal-) resolution seeing-free observations in a 1.1~\AA\ wide passband covering the line core of Ca~{\sc ii}~H 3969~\AA\ from the second flight of the {\sc Sunrise} balloon-borne solar observatory. The entire field of view, spanning the polarity inversion line of an active region close to the solar disk center, is covered with bright, thin, and very dynamic fine structures.
Our analysis reveals the prevalence of transverse waves in SCFs with median amplitudes and periods on the order of $2.4\pm0.8$~km/s and $83\pm29$~s, respectively (with standard deviations given as uncertainties).
We find that the transverse waves often propagate along (parts of) the SCFs with median phase speeds of $9\pm14$~km/s. While the propagation is only in one direction along the axis in some of the SCFs, propagating waves in both directions, as well as standing waves are also observed.
The transverse oscillations are likely Alfv\'{e}nic and are thought to be representative of magnetohydrodynamic kink waves.
The wave propagation suggests that the rapid high-frequency transverse waves, often produced in the lower photosphere, can penetrate into the chromosphere with an estimated energy flux of $\approx15$~kW/m$^2$.
Characteristics of these waves differ from those reported for other fibrillar structures, which, however, were observed mainly in the upper solar chromosphere.

\vspace{1mm}
\end{abstract}


\section{Introduction}
\label{sec:intro}

The solar chromosphere is a highly structured environment exhibiting elongated features at different spatial scales~\citep{Judge2006,Rutten2007,Woger2009}. Depending on their physical and dynamical properties, these thread-like structures have been given names such as fibrils, dynamic fibrils, mottles, straws, rosettes, spicules, or rapid blue/redshifted events (RBEs/RREs) in the literature~\citep{DePontieu2007c,Rutten2007,Rouppe2009}; see \citet{Rutten2012}; see also \citet{Tsiropoula2012} for a recent review. Most of these structures often share some properties and appear to be associated with photospheric concentrations of the magnetic field such as magnetic bright points and plages. These features have been mostly observed in H$\alpha$~6563~\AA\ and Ca~{\sc ii}~8542~\AA\ images (representing the mid-to-upper chromosphere) on the solar disk~\citep{Vecchio2007,Cauzzi2008}, and in Ca~{\sc ii}~8542~\AA,\ Ca~{\sc ii}~H/K, and H$\alpha$ images off the limb (e.g., \citealt{Pasachoff2009,Pereira2016}; see~\citealt{Pereira2014}).

In particular, internetwork fibrils (usually seen in H$\alpha$ images) are relatively long and are often considered to outline the magnetic field lines in the chromosphere. They appear to constitute magnetic canopies in the mid-to-upper chromosphere, along which flows may reach to the transition region (with some signatures observed in, e.g., He~{\sc ii}~304~\AA;~\citealt{Rutten2012}). Their correspondence to the chromospheric magnetic field topology has been partly confirmed by comparisons with full Stokes observations in Ca~{\sc ii}~8542~\AA\ \citep{de-laCruz2011}. These elongated features are, however, shorter and more dynamic in and close to network and active regions and are sometimes seen extending into the upper atmosphere~\citep{Suematsu1995}. The most dynamic of the active region elongated features are jet-like features seen in, e.g., Ly~$\alpha$ observations~\citep{Patsourakos2007,Koza2009}.

From observations in a 0.3~\AA\ wide Ca~{\sc ii}~K passband, \citet{Zirin1974} claimed the existence of Ca~{\sc ii}~K bright fibrils matching dark fibrils in H$\alpha$ filtergrams. Lifetime and length of those fibrils was thought to depend on the inclination of the field lines \citep{Marsh1976}.

Investigating the on-disk Ca~{\sc ii}~H/K bright fibrils (in the lower chromospheric layers) has remained a challenging task.
First, these very thin and bright features can only be identified in relatively narrowband Ca~{\sc ii}~H/K filters~\citep{Reardon2007,Beck2013}. Thus, broader filters, such as that of \textit{Hinode}/SOT (\citealt{Tsuneta2008}; FWHM of 3.0~\AA), sample a wide range in formation height, from the mid photosphere (in their wings), at around the temperature minimum at the H$_1$/K$_1$ dip, reaching up to the mid chromosphere at the H$_2$/K$_2$ emission peaks around their line-cores~\citep{Vernazza1981}. Hence, the \textit{Hinode}/SOT  Ca~{\sc ii}~H filter is too broad to see the slender bright fibrils in the lower chromosphere. In addition, precise detection of the very thin and dynamic Ca~{\sc ii}~H/K fibrils needs data with high spatial and temporal resolution~\citep{Pietarila2009}. 

Second, ground-based observations of temporal evolution of these fine structures with, e.g., a narrowband Ca~{\sc ii}~H filter (FWHM $\approx1.0$~\AA)\ of the Swedish Solar Telescope (SST;~\citealt{Scharmer2003}), need exceptional seeing conditions for a relatively long duration.

The only thorough study of such slender bright fibrils, to our knowledge, was provided by \citet{Pietarila2009} who demonstrated that these thin structures are ubiquitous in high spatial resolution and high-quality time series of images of a small active region recorded in a 1.5~\AA\ Ca~{\sc ii}~K filter with the SST. We note that fibril-like structures (often dark) above sunspots have been also reported in observations of the SST narrowband Ca~{\sc ii}~H filter~\citep{Henriques2013,Henriques2015}.

A range of waves and oscillations that are thought to be important for heating the upper solar atmosphere have been observed in these fibrillar structures (e.g.,~\citealt{Lin2007,Pietarila2011}; also \citealt{Tsiropoula2012} for a review). These waves are often produced in the photosphere by, e.g., the buffeting of footpoints of their underlying magnetic elements, and propagate (longitudinally or transversely) along the magnetic field lines on different scales. The fibrils appear to act as a guide for the propagation of the MHD waves~\citep{Hansteen2006,DePontieu2007a} whose leakage to the upper solar atmosphere depends on, e.g., the inclination angle of the fibrils~\citep{Michalitsanos1973,Bel1977}. Thus, dissipation of energy carried by these MHD waves (of different types) is of interest for the energy budget at different heights in the chromosphere and in the corona (\citealt{DePontieu2011}, also see \citealt{Jess2015b} for a recent review).

In particular, transverse waves (in both standing and propagating states) have been observed in a number of features throughout the solar atmosphere, such as in magnetic elements in the lower solar atmosphere \citep[e.g.,][]{Stangalini2015,Jafarzadeh2017a}, in spicules and filaments in the upper chromosphere \citep[e.g.,][]{DePontieu2007,Lin2007}, and in prominences and loops in the corona \citep[e.g.,][]{Aschwanden1999,Okamoto2007,Tomczyk2007}. These transverse waves have often been interpreted as MHD kink and/or Alfv\'{e}n waves \citep{Spruit1982,Nakariakov2005,Erdelyi2007b}. To date, however, these transverse waves have not been fully quantified in the lower solar chromosphere (e.g., in Ca~{\sc ii}~H/K fibrils).

\begin{figure*}[!tp]
\centering
    \includegraphics[width=18cm, trim = 0 0 0 0, clip]{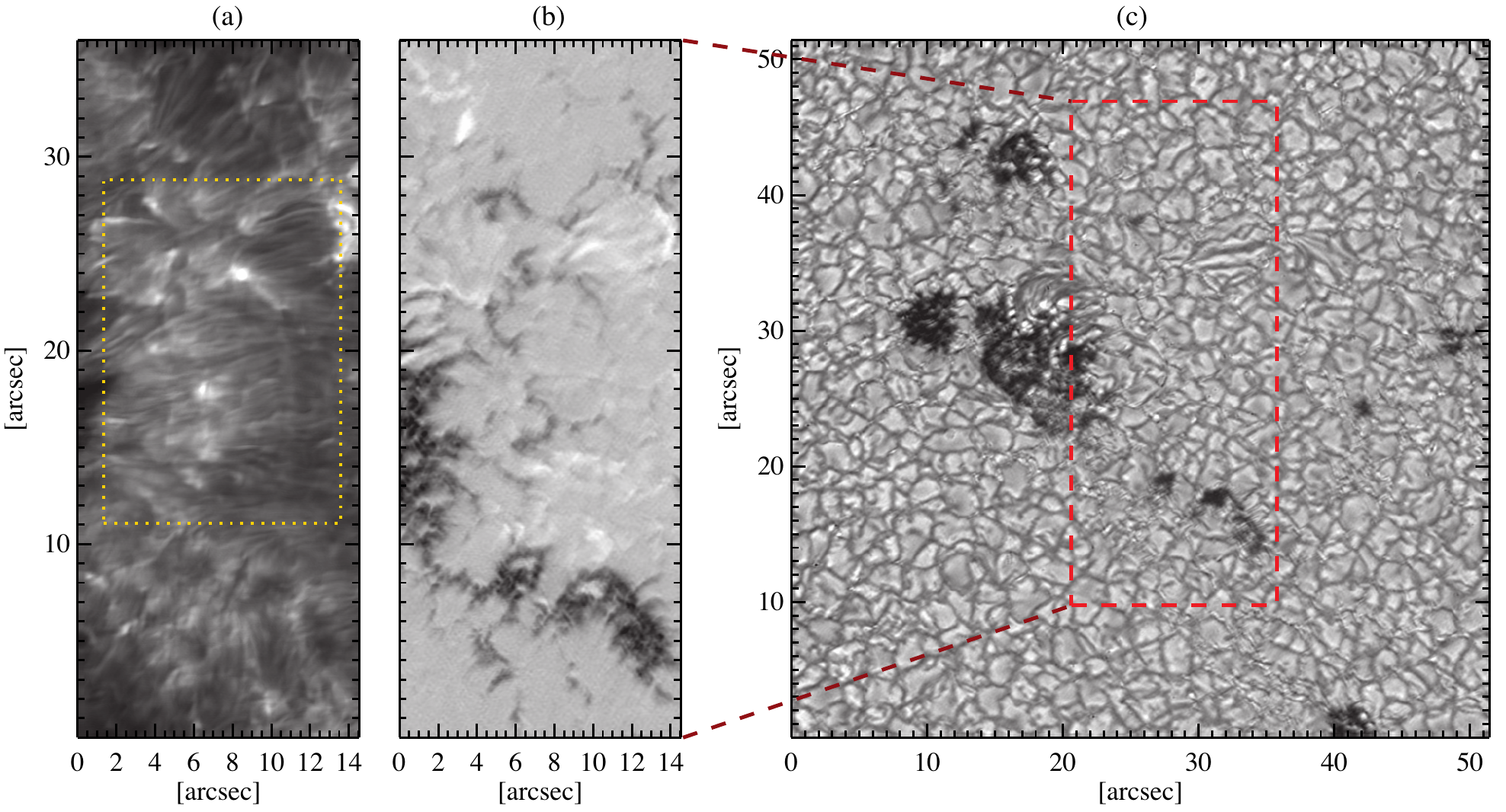}
  \caption{Examples of the {\sc Sunrise}/SuFI Ca~{\sc ii}~H narrowband image (a), the co-temporal, co-aligned Stokes $V$ image from {\sc Sunrise}/IMaX cropped to the SuFI field of view (b), and the corresponding Stokes $I$ continuum image at its original size and resolution (c). The red dashed-line box in panel (c) indicates the same field of view as those in panels (a) and (b). The yellow box in panel (a) marks a region of interest for which results of image restorations are illustrated in Figure~\ref{fig:segment}.}
  \label{fig:obs}
\end{figure*}

In this paper, we present ubiquitous transverse oscillations in slender Ca~{\sc ii}~H fibrils (SCFs) from a high-quality and relatively long time series of images recorded at high spatial and temporal resolution (in a 1.1~\AA\ narrowband filter) with the {\sc Sunrise} balloon-borne solar observatory. We quantify the oscillations by analyzing the swaying motions in several spatial locations along their axes in the course of their lifetimes. We provide parameters of the oscillations/waves and suggest the role they may play in the localized heating of the lower solar chromosphere.

\vspace{1mm}

\section{Observations}
\label{sec:obs}

Our analysis is based on high-resolution images in the Ca~{\sc ii}~H narrowband filter (FWHM$\approx1.1$~\AA) acquired with the {\sc Sunrise} Filter Imager (SuFI;~\citealt{Gandorfer2011}) aboard the 1-meter {\sc Sunrise} balloon-borne solar observatory~\citep{Solanki2010,Barthol2011,Berkefeld2011} in June 2013, i.e., its second flight (see \citealt{Solanki2017} for more details on the second flight of {\sc Sunrise}). The seeing-free, uninterrupted time series of images were recorded for $\approx60$~minutes (between 23:39~UT on 2013 June 12 and 00:38~UT on 2013 June 13) with a cadence of 7~s. The Ca~{\sc ii}~H filtergrams cover a field of view (FOV) of ($15\times38$)~arcsec$^2$, after image restoration~\citep{Hirzberger2010,Hirzberger2011}, in the immediate vicinity of an active region close to the solar disk center (with the cosine of heliocentric angle $\mu \approx 0.93$). The images include a few small pores and plages in their FOVs. These high-quality data were further corrected for wavefront aberrations by means of multi-frame blind deconvolution (MFBD;~\citealt{vanNoort2005}).
Finally, the time series of images were corrected for field rotation by determining the rotation angle from the telescope pointing model (which depends on the position of the Sun on the sky). Temporal misalignments were also corrected for, which included determining the geometrical shift and scale of each frame with respect to its preceding image, using a cross-correlation technique. There could still, however, remain residual alignment errors between consecutive frames due to, e.g., some jitter from the telescope. Following \citet{Shine1994}, we employed a de-stretching approach to remove the variable distortion of the images. In short, each frame was divided into a grid of overlapping sub-images whose spatial misalignments with their corresponding ones in the following image were determined and corrected.

We note that in addition to the line core (sampling the low-to-mid chromospheric heights), the 1.1~\AA\ Ca~{\sc ii}~H passband also includes a small portion of the inner line wings contaminated by photospheric radiation. However, in the presence of strong magnetic fields, the contribution of the line core (through further enhancement of the emission peaks) is much larger than that of the wings~\citep{Beck2013}. Hence, our images mainly represent chromospheric heights.

Additionally, the {\sc Sunrise} Imaging Magnetograph eXperiment (IMaX;~\citealt{Martinez-Pillet2011}) undertook simultaneous full Stokes ($I$, $Q$, $U$, and $V$) observations of the magnetically sensitive Fe~{\sc i}~$5250.2$~\AA\ line, which overlap with the first $\approx17$~minutes of our Ca~{\sc ii}~H observations. The IMaX images cover a larger FOV of ($51\times51$)~arcsec$^2$ and are used to identify the magnetic areas based on Stokes $V$ observations.

Figure~\ref{fig:obs}(a) illustrates an example of the SuFI Ca~{\sc ii}~H image in a selected time stamp. Its co-spatial and co-temporal Stokes $V$ image in the IMaX Fe~{\sc i}~$5250.2$~\AA,\ (an average over four wavelength positions at $\pm80$ and $\pm40$~m\AA\ from the line center), as well as its co-temporal Stokes $I$ continuum frame (but with the full FOV of IMaX), are also plotted in panels (b) and (c), respectively. The FOV of the images in panels (a) and (b) is marked with the red dashed box on the Stokes $I$ continuum image. The yellow box in panel (a) includes the region of interest (ROI) whose analysis is discussed in Section~\ref{sec:analysis}. The ROI is chosen to include a region with a high density of fibrils, but excluding the pores. We note that the FOV of the images in Figure~\ref{fig:obs} is vertically flipped and slightly rotated with respect to the true orientation on the Sun. For the correct orientation, see \citet{Solanki2017}.

\vspace{1mm}
\section{Analysis and Results}
\label{sec:analysis}

The study of individual SCFs embedded in the spatially dense forest of bright, extremely thin, and highly dynamic fibrillar structures observed in the SuFI narrowband Ca~{\sc ii}~H image sequences needs several steps of image processing prior to the actual analysis. It is specially important since, in addition to these bright structures, Ca~{\sc ii}~H images also exhibit extended bright background and/or other brightenings due to, e.g., shock waves and/or reversed granulation. We note that the latter has, however, less impact on our images, whose observations in a relatively narrowband filter within an active region sample higher chromospheric layers (with less imprint of the reversed granulation) compared to those in the quiet Sun (see, e.g., \citealt{Riethmuller2013c}). 

After the image processing (described in the following subsection), the SCFs are then segmented using a semi-automated procedure. This is followed by tracing the detected SCFs in the times series of images and investigating possible transverse oscillations along their axes. These oscillations are further quantified by determining their amplitudes, periods, and transverse velocities. Phase speeds of propagating waves are also calculated, when possible.

\vspace{1mm}

\subsection{Image Restoration and Fibril Identification}
\label{subsec:restoration}

In order to facilitate the identification of the bright SCFs in extended bright regions sampled with the Ca~{\sc ii}~H passband, we restored the images based on a process described by \citet{Jafarzadeh2013a}. For discrete images, this approach minimizes inhomogeneities due to non-uniform background solar intensity, noise, and geometric distortion produced by slightly rectangular pixels. The image restoration employs a real-space spatial bandpass filter to eliminate localized extended brightenings (i.e., low spatial frequencies) caused by, e.g., shock waves, followed by corrections for high-frequency noise.

The localized and elongated features in the resulting images possess enhanced intensity profiles that are proportional to their original brightness. These include, however, small structures that are generally not SCFs (e.g., H$_{2V}$ grains or magnetic bright points; see \citealt{Carlsson1997} and \citealt{Jafarzadeh2013a}) and/or spurious artifacts introduced by the image restoration. Such features are discarded in the next steps.

\begin{figure*}[!tp]
\centering
    \includegraphics[width=18cm, trim = 0 0 0 0, clip]{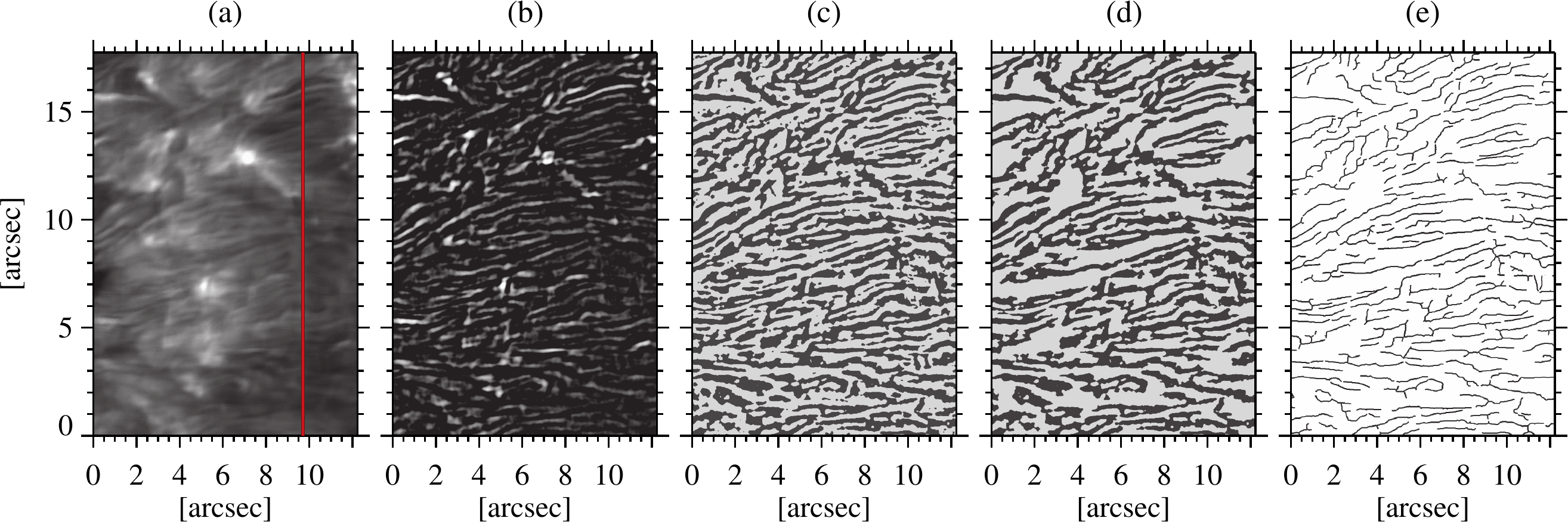}
  \caption{Segmentation and modeling of slender Ca~{\sc ii}~H fibrils (SCFs) in the region of interest indicated by the yellow box in Figure~\ref{fig:obs}(a). (a) Original intensity image. (b) Restored image, where the non-uniform, extended background intensity of the original image has been subtracted. The brightness is enhanced for better visibility. (c) Inverted binary image of panel (b). (d) Same as panel (c), except that particularly small and short features have been removed from the image. (e) Skeleton of the features shown in panel (d). The red line in panel (a) marks a cut through the image, along which displacement oscillations are studied (see Figure~\ref{fig:time_distance}).}
  \label{fig:segment}
\end{figure*}

An example of a Ca~{\sc ii}~H image for the ROI (marked in Figure~\ref{fig:obs}(a) with a yellow box) and its restored image are shown in Figures~\ref{fig:segment}(a) and \ref{fig:segment}(b), respectively.

The small, bright features that are not of interest for the present study are then detected and removed from each restored image. This is implemented using a modified ``blob analyzer'' algorithm, where all isolated elongated features in a bi-level image (i.e., a binary of the restored image) are automatically identified. All detected features are given a unique identity and their areas as well as their (major and minor) axes are determined by fitting ellipses to the features. Structures with major axes shorter than 2~arcsec are excluded from the images. Illustrated in Figures~\ref{fig:segment}(c) and \ref{fig:segment}(d) are the inverted binary version of the restored image shown in Figure~\ref{fig:segment}(b), before and after removal of small features, respectively. We refer to the latter as the ``clean image'' in the following. Furthermore, spines of the elongated features are depicted in Figure~\ref{fig:segment}(e). These skeletons are found as the median axes of the features in Figure~\ref{fig:segment}(d) and represent the overall shape of the SCFs regardless of their thickness. It turned out that the majority of the SCFs, as evidenced by the skeletons in Figure~\ref{fig:segment}(e), are oriented along the $x$-axis (sometimes with a small angle).

Identifying SCFs from the clean images still needs careful human intervention. This is particularly important since there are features composed of several bright events (including individual fibrils) superposed on a small area. Thus, the time series of clean images are visually inspected. Features meeting the following criteria are manually selected as our SCF candidates. (1) They can be visually observed in at least three consecutive frames (to avoid false detections as a result of, e.g., artifacts). (2) They are, to a large extent, isolated, i.e., their edges are clearly seen in the course of their lifetimes and no other significant brightenings occur at their spatial locations. In this manner interacting fibrils are excluded.

The SCFs appear not to move over a large distance during their relatively short apparent lifetimes, which are likely lower limits of their true lifetimes (see \citealt{Gafeira2017a}). Instead, they sway, more or less, around the same location in the time series of images. We use the CRISPEX tool \citep{Vissers2012} for the visual inspections; since it provides not only the ease of browsing image sequences, but also space-time plots for selected spatial locations, with different lengths and orientations. The latter is particularly convenient for identifying the isolated SCFs.

\vspace{1mm}

\subsection{Transverse Oscillations}
\label{subsec:transoscillations}

\subsubsection{Ubiquity}
\label{subsec:ubiquity}

\begin{figure}[!h]
\centering
    \includegraphics[width=8.4cm, trim = 0 0 0 0, clip]{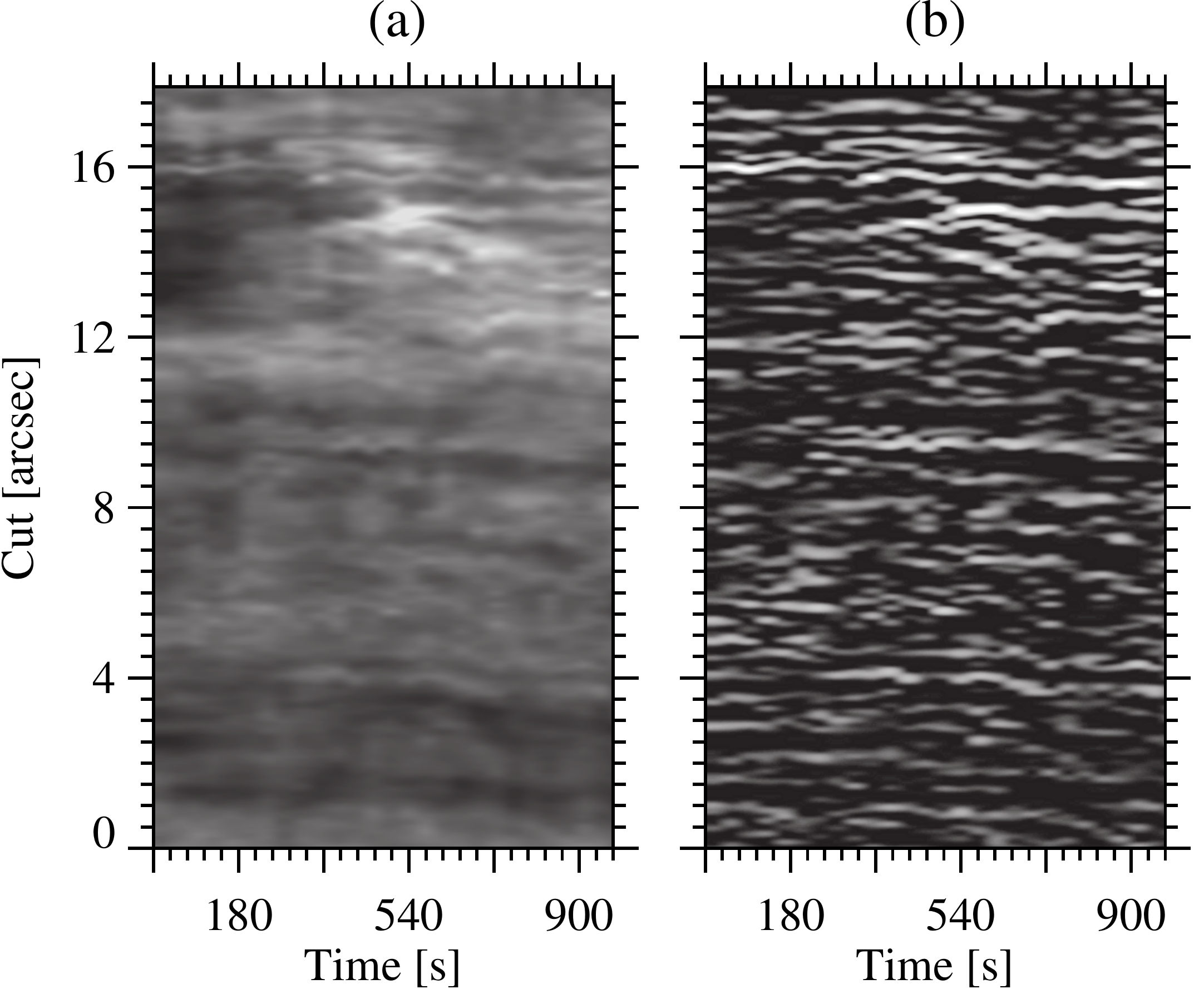}
  \caption{Illustration of ubiquity of transverse oscillations in the slender Ca~{\sc ii}~H fibrils (SCFs). (a) A space-time plot along a cut depicted with the red line in Figure~\ref{fig:segment}(a). (b) Same as (a), but obtained from the corresponding restored images, with the brightness enhanced for better visibility.}
  \label{fig:time_distance}
\end{figure}

To inspect transverse oscillations in the SCFs, we create artificial slits oriented more or less perpendicular to the SCFs axes at any given time. Spacetime plots are then extracted by stacking the slits at every time step (i.e., every 7~s). These plots show the intensity variations with time along the slits. Figure~\ref{fig:time_distance}(a) illustrates a space-time plot of the SCFs along a cut perpendicular to the axes of the majority of the linear features in our ROI, following the red line in Figure~\ref{fig:segment}(a). It clearly reveals transverse oscillations, i.e. displacements, normal to the spines of the SCFs. Because of the substantial superposition of various bright features in the Ca~{\sc ii}~H images, the ubiquitous transverse motions are best seen in a space-time plot from the restored/clean images in which the background solar intensity and noise are diminished (Figure~\ref{fig:time_distance}(b)). As noted in Section~\ref{subsec:restoration}, most of the SCFs in our ROI (Figure~\ref{fig:segment}(a)) are along the $x$-axis, thus, the cut has been chosen to be in $y$ direction for the example shown in Figure~\ref{fig:time_distance}. For simplicity, and better visibility of the wave patterns, the variations are plotted for the first quarter of the time series of images in our dataset.

While the majority of the identified SCFs can be followed only for a relatively short time (i.e., shorter than $\approx100$~s), events that are consistently visible longer undergo periodic oscillations normal to the SCFs' spines. These may represent transverse waves in the fibrils. In addition, we can also note the presence of some static bright features, which may represent non-oscillatory SCFs (or SCFs with small-amplitude oscillations that are not detected). We note that those brightenings may, however, also represent other non-SCF events.

\subsubsection{Characteristics}
\label{subsec:characteristics}

We determine period, displacement, and transverse velocity of identified oscillations from the space-time diagrams. The oscillations are visually selected in a two-step procedure: (1) wave-like patterns are pre-identified from space-time plots, similar to those shown in Figure~\ref{fig:time_distance}, but along cuts perpendicular to individual detected fibrils. (2) Fibrils corresponding to the wave patterns are visually inspected in the time series of images (from both original and restored/clean images) in a small area around the SCFs. The latter confirms whether the wave-like patterns, identified in the first step, truly correspond to transverse oscillations in isolated SCFs, i.e., if they appear to sway in the image sequence played as a movie.

The selected wave patterns are then quantified using a semi-automated procedure where spatial locations of the oscillations are determined (see, e.g, \citealt{Aschwanden2002,Verwichte2004,Morton2014b} for similar techniques). Several points are visually selected on the approximate centroid of each oscillation at fixed cadence.
This facilitates automatic measurements of the precise location of the spine by fitting a Gaussian at the selected points, perpendicular to the axis of the wave pattern in the space-time diagram. The Gaussian fits provide us with the locations and widths of the SCF along the cut at different times. The locations were determined using the same algorithm as introduced by \citet{Jafarzadeh2013a} which technically has a precision of 0.05~pixel (better than 1~km). However, in accordance with a discussion by \citet{Jafarzadeh2013a} and \citet{Jafarzadeh2014b}, we consider a more conservative uncertainty of 0.5~pixel ($\approx7$~km).

\begin{figure}[!tp]
\centering
    \includegraphics[width=8.4cm, trim = 0 0 0 0, clip]{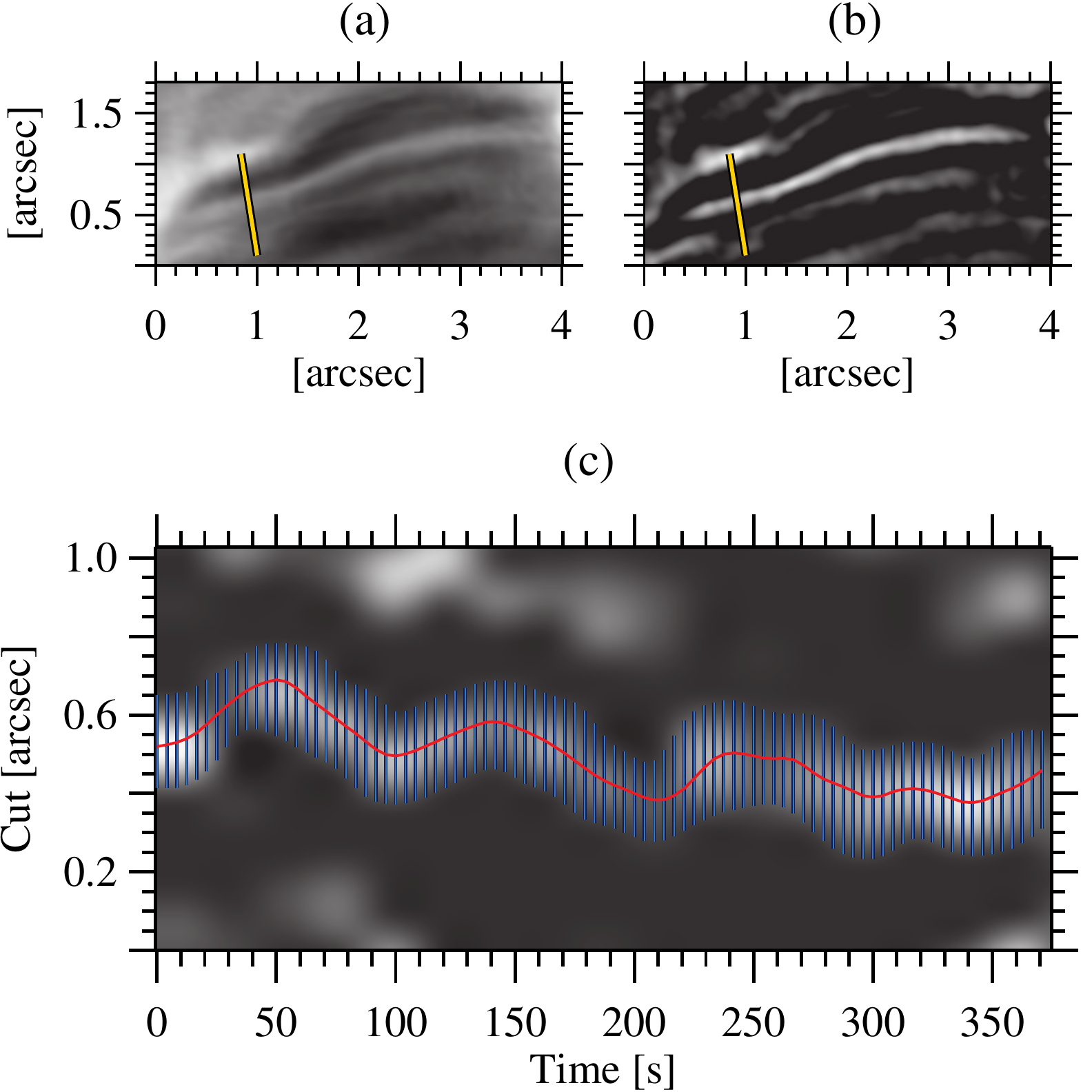}
  \caption{Original (a) and restored (b) images of a slender Ca~{\sc ii}~H fibril (SCF) whose transverse oscillations along a cut perpendicular to its long axis is illustrated in panel (c). The yellow lines on panels (a) and (b) show the cut. Centroid of the transverse oscillation (the red line; smoothed by a box average of 0.05~arcsec width in space and 5~s width in time) and the width of the SCF (the blue vertical lines), determined using a semi-automated procedure (see the main text), are also depicted on the space-time plot.}
  \label{fig:detect_example}
\end{figure}

Figures~\ref{fig:detect_example}(a) and \ref{fig:detect_example}(b) represent a small area including a sample SCF (i.e., in the original and restored images, respectively). Transverse oscillation of the SCF along the cut marked in yellow in panels (a) and (b) is shown in panel (c). Width (the vertical blue lines) and the centroid of the oscillation (the red line; smoothed by a box average of 0.05~arcsec width in space and 5~s width in time), determined from a Gaussian fit at the equally spaced locations along the SCF, are also illustrated.

We note that the measurements are only implemented on oscillations lasting at least one full period. Prior to the analysis of each oscillation, we detrend the wave pattern to remove linear (systematic) drift. Then, following \citet{Leenaarts2015}, periods ($P$), transverse displacements ($D$), and transverse velocities ($v$) are calculated as

\begin{equation}
	P_{i} = t_{i+1}- t_{i-1}\,,
	\label{equ:period}
\end{equation}

\begin{equation}
	D_{i} = \frac{\left |x_{i+1}-x_{i}   \right |}{2}\,,
	\label{equ:displ}
\end{equation}

\begin{equation}
	v_{i} = \frac{\pi\, \left ( D_{i-1}+D_{i} \right )}{P_{i}}\,,
	\label{equ:transvel}
\end{equation} 
\noindent
where $x$ and $t$, respectively, represent the space and time coordinates of the $i$th extremum of the oscillation, whereby $x$ is chosen to lie perpendicular to the time-averaged spine of the SCF.

\vspace{-2mm}
\subsection{Wave Analysis}
\label{subsec:wave}

The majority of identified SCFs are uninterruptedly visible for a relatively short time, hence in most of the detected features the oscillatory behavior cannot be traced. Thus, wave propagation is investigated in only a small fraction of the oscillations identified in Section~\ref{subsec:characteristics}. 

Propagation of transverse disturbances along a SCF can be deduced when oscillations are studied in at least two spatially separate cuts across the fibril axis. Propagation speeds in the SCF can then be estimated from phase differences between oscillatory motions at the different positions.

\begin{figure}[!htp]
\centering
    \includegraphics[width=8.2cm, trim = 0 0 0 0, clip]{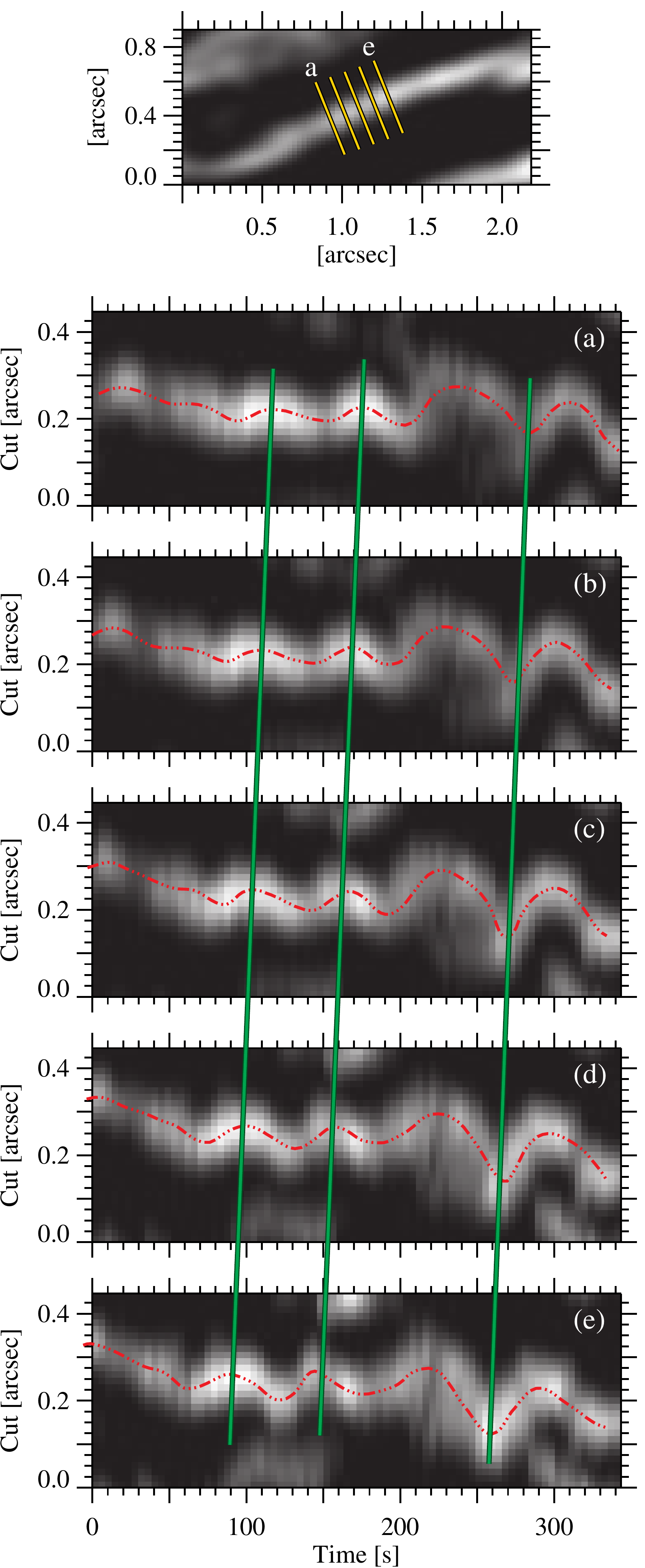}
  \caption{Phase speeds in an example slender Ca~{\sc ii}~H fibril (SCF) with an indication of wave propagation. The curves in panels (a)-(e) represent displacement of the SCF along a series of cuts across the fibril shown in the upper panel, from left to right, respectively. The red triple-dot--dashed lines are the centroids of a Gaussian fit to the oscillations smoothed by convolving with a boxcar filter of 0.05~arcsec width. The green lines indicate waves propagation in the SCF in a direction corresponding to from right to left in the top panel.}
  \label{fig:wave_propagate}
\end{figure}

Figure~\ref{fig:wave_propagate} shows an example of a relatively long-lived SCF (top panel) along with transverse oscillations (panels (a)-(e)) in the five cuts marked in yellow in the top panel (from left to right, respectively). We note that the cuts are chosen to lie relatively close to each other, where the SCF exists over the longest period of time (since the length of the SCF varies with time; see \citealt{Gafeira2017b}). The red triple-dot--dashed lines illustrate centroids of the oscillations. The green solid lines indicate linear fits to three selected extrema. These clearly demonstrate wave propagation in the SCF in one direction (i.e., from right to left along the SCF shown in the top panel). The transverse oscillations shown in Figures~\ref{fig:wave_propagate}(a)--(e) include periods, transverse displacements, and velocity amplitudes in the range of 45-105~s, 15-70~km, and 1.5-4~km/s, respectively. 

\begin{figure}[!htp]
	\centering
	\includegraphics[width=8.2cm, trim = 0 0 0 0, clip]{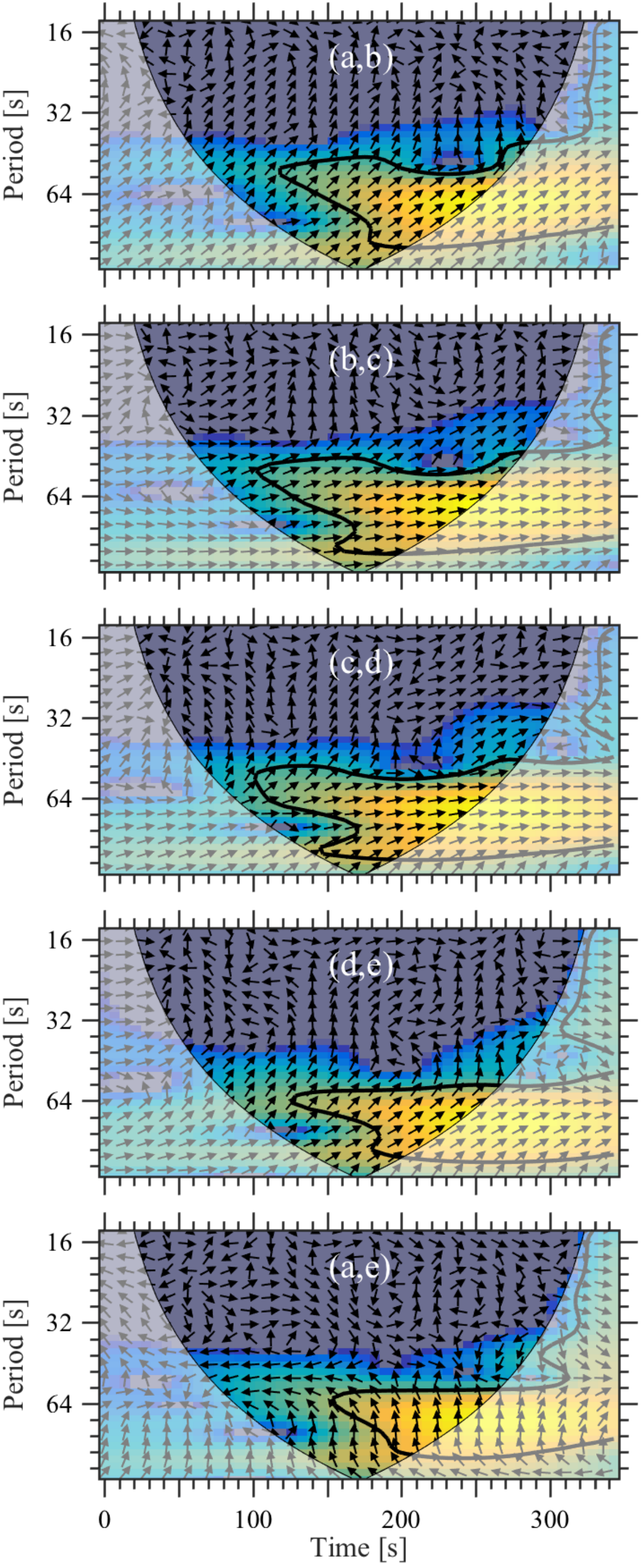}
	\caption{Wavelet cross-power spectra between pairs of transverse perturbations shown in Figure~\ref{fig:wave_propagate} (the pairs are indicated near the top of each panel). The background colors represent power spectra (with the highest power being colored yellow and the lowest power gray). The $5\%$ significance level against red noise is marked with a thick solid contour in each panel. The arrows depict the relative phase relationship between the pair of oscillations (with in-phase pointing right, anti-phase pointing left, and the second oscillation leading the first one by $90^\circ$ pointing straight down). The bleached areas indicate the cone of influence (see the main text).
	 }
	\label{fig:xrt}
\end{figure}

We use wavelet analysis \citep{Daubechies1990,Torrence1998} to characterize wave propagation in the SCFs. In the method used here, which is described in detail by \citet{Jafarzadeh2017a}, the following steps are taken. (1) Spectral power of each transverse oscillation is simultaneously computed in the time and frequency domains. (2) Cross wavelet transforms, representing correlations between the wavelet power spectra of the oscillations at any two locations in a SCF (i.e., from the cuts across the fibrils) are determined. (3) Phase differences between each pair of oscillations whose correlated power spectra lie significantly above the noise are calculated. A significance level of $5\%$ is considered for the latter cross-power spectra.

\begin{figure}[!htp]
\centering
    \includegraphics[width=8.2cm, trim = 0 0 0 0, clip]{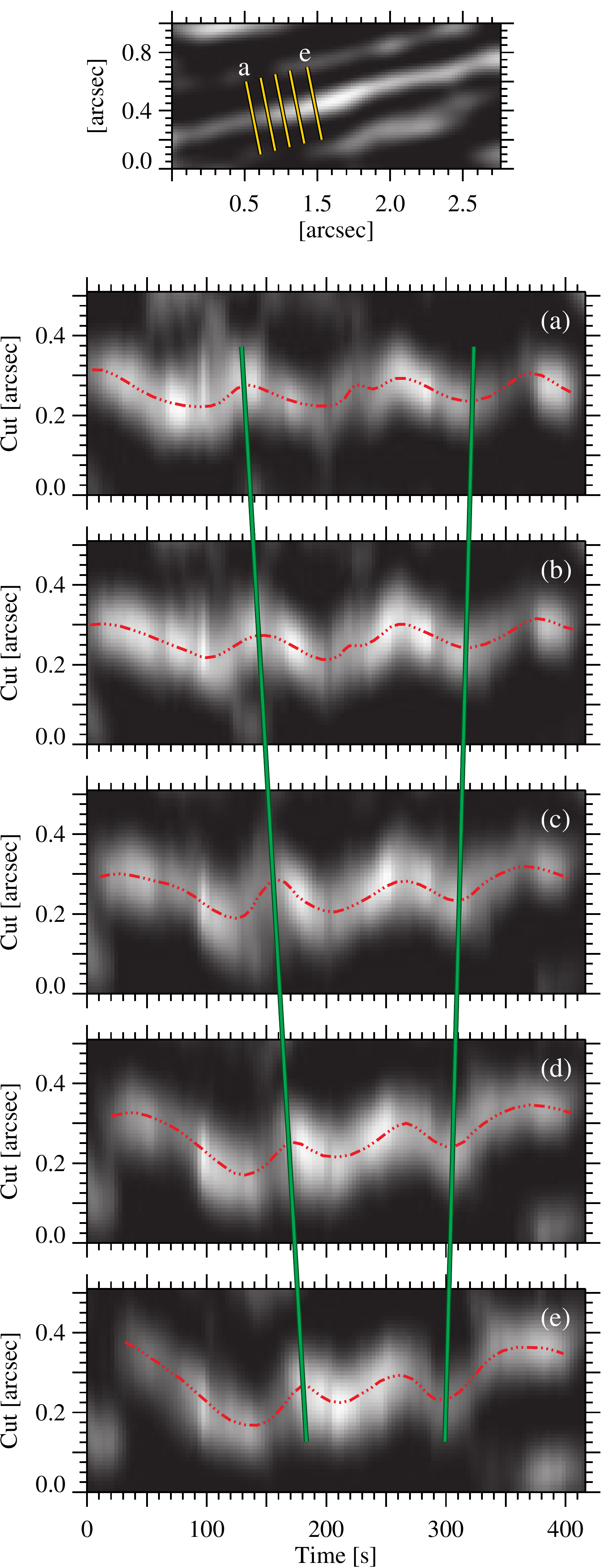}
  \caption{Same as Figure~\ref{fig:wave_propagate}, but for a complex example. The green lines highlight waves in a SCF propagating in opposite directions.}
  \label{fig:wave_propagate2}
\end{figure}

\begin{figure}[!htp]
	\centering
	\includegraphics[width=8.2cm, trim = 0 0 0 0, clip]{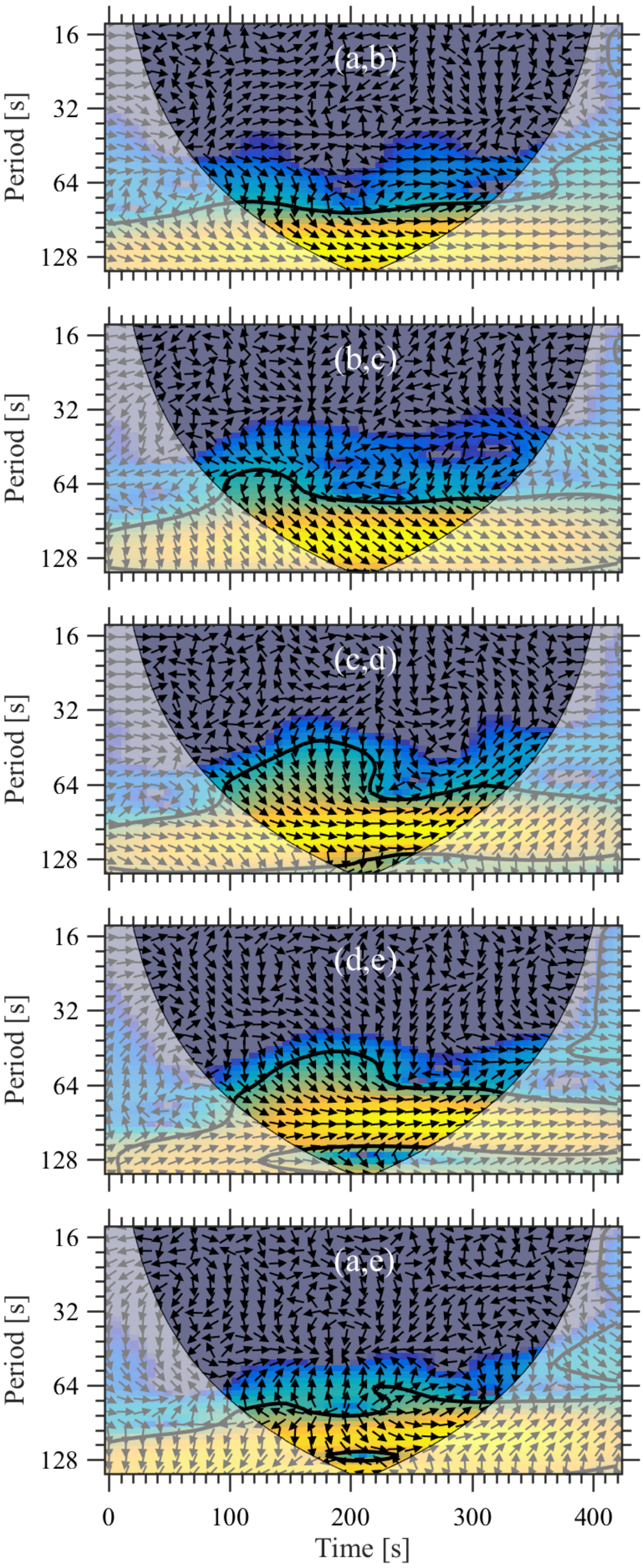}
	\caption{Same as Figure~\ref{fig:xrt}, but for transverse oscillations shown in Figure~\ref{fig:wave_propagate2}.
	 }
	\label{fig:xrt2}
\end{figure}

Figure~\ref{fig:xrt} represents wavelet cross-power spectra between oscillations at pairs of locations (a-e) along the SCF shown in Figure~\ref{fig:wave_propagate}. The corresponding pairs of locations are indicated near the top of each panel. The power is represented by the background color, with yellow marking the highest power and gray the lowest. Regions with bleached color mark areas that are subject to the edge effect (the so-called cone of influence, COI). Thus, only values outside the COI are analyzed.
Phase lags between oscillation pairs are depicted by the black arrows in selected locations.
Arrows pointing right represent in-phase oscillations, i.e., $\varphi=0$, (and anti-phase oscillations when pointing left) and those pointing down indicate that the second oscillation leads the first one by $90^\circ$, i.e., a wave propagation from left to right in the SCF illustrated in the top panel of Figure~\ref{fig:wave_propagate}. The thick solid contours mark the $5\%$ significance level against red noise (increasing power with decreasing frequency), within which and outside the COI (i.e., in the region of consideration) the values of phase angles ($\varphi$) along with their corresponding periods ($P$) result in time differences ($\tau$) that the waves need to travel between the pair of locations:

\begin{equation}
	\tau = \frac{\varphi P}{2\pi}\,.
	\label{equ:phase_time}
\end{equation}

From the direction of the arrows in the regions of consideration in Figure~\ref{fig:xrt}, a range of phase angles of the propagating waves in one direction (as was also noted from Figure~\ref{fig:wave_propagate}) are found in different times and locations in the sample SCF.

Knowing the distance between the cuts along which the oscillation pairs were measured as well as their corresponding time lags (computed from Equation~(\ref{equ:phase_time})) we calculate phase speeds of the waves, representing the speed of wave propagation along a SCF (or a part of it).
A median phase speed of $11\pm9$~km/s was obtained for the waves propagating in the example SCF studied here (see Figures~\ref{fig:wave_propagate} and \ref{fig:xrt}). This large uncertainty in the propagation speeds obtained in a single SCF is likely due to the small distance between the individual cuts, so that even non-linearities in the wave-form or similar effects can result in rather different wave speeds. We note that the median phase speed determined from the wavelet analysis is in agreement with that calculated directly from shifts of the wave patterns in Figure~\ref{fig:wave_propagate}. This sample SCF was found to have a median period, transverse displacement, and velocity amplitude of $64\pm14$~s, $28\pm20$~km, and $1.8\pm1.1$~km/s, respectively.

While propagation in only one direction is the most common case in our sample (it is seen in $\approx70\%$ of the investigated SCFs), propagating waves in opposite directions are also observed in some of the SCFs under study. Figures~\ref{fig:wave_propagate2} and \ref{fig:xrt2} represent one example of such a complex wave propagation. The green lines in the former figure connect two selected extrema of the oscillations (along the five cuts shown in the top panel) and clearly demonstrate an oppositely propagating wave in the SCF: the left bump has a positive phase-lag, whereas the right dip has a negative phase relationship. From the direction of the arrows in the regions of consideration in Figure~\ref{fig:xrt2}, a range of phase angles of the propagating waves in opposite directions as well as in-phase/anti-phase oscillations are found in different times and locations in the sample SCF. The phase differences results in median propagating speeds of $15\pm12$~km/s and $7\pm5$~km/s for the oppositely moving waves in this SCF.

\subsection{Statistics}
\label{subsec:statistics}

\begin{figure*}[!tph]
\centering
    \includegraphics[width=17cm, trim = 0 0 0 0, clip]{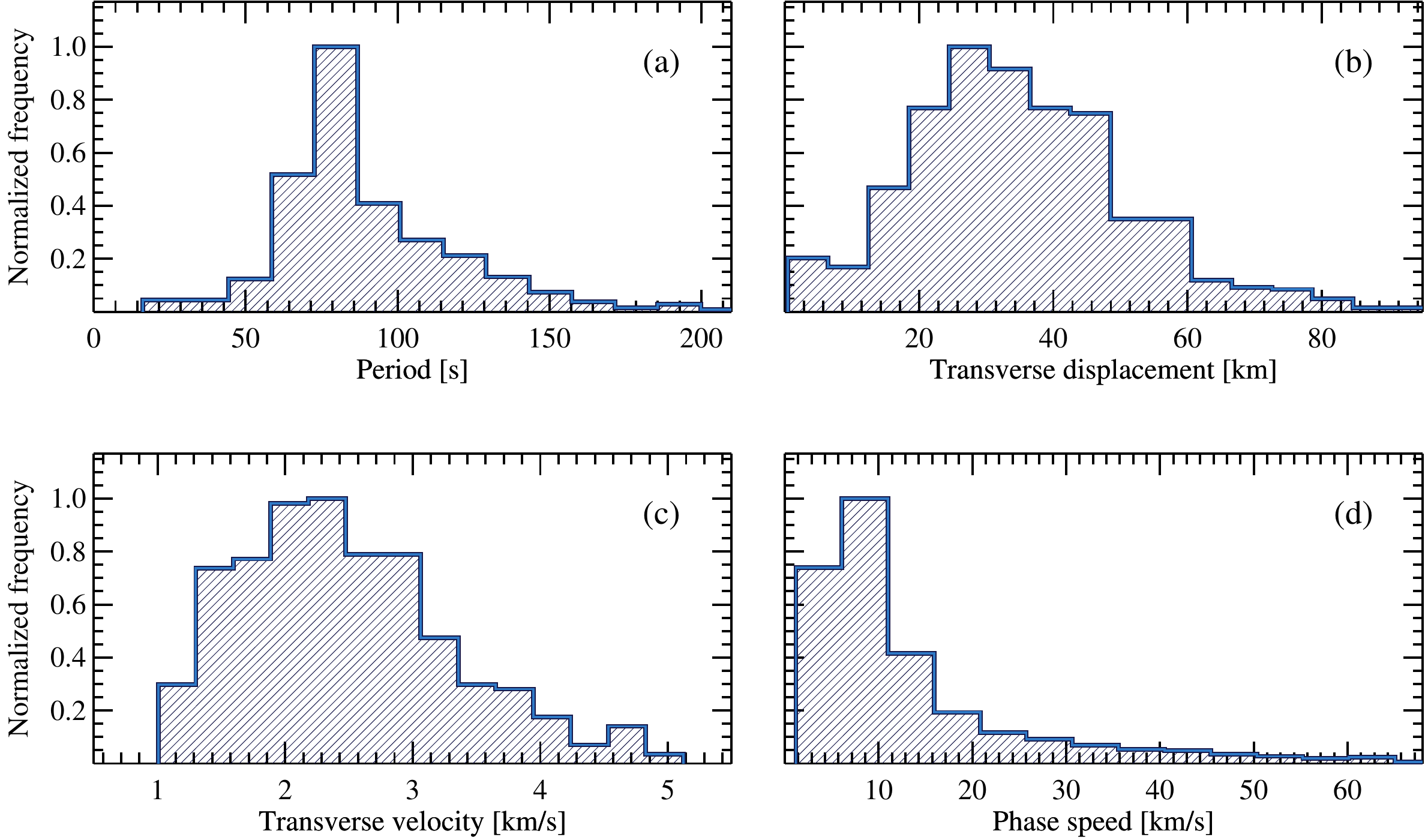}
  \caption{Distributions of period (a), transverse displacement (b), transverse velocity (c), and phase speed (d) of oscillations in slender Ca~{\sc ii}~H fibrils. The histograms are normalized to their maximum values.}
  \label{fig:statistics}
\end{figure*}

We characterized transverse oscillations in a total of 134 SCFs observed with {\sc Sunrise}/SuFI. This includes measurements of their periods, transverse displacements, and velocity amplitudes (transverse velocities). Among all of these, it was possible to trace waves (i.e., characterize wave propagation) in only 23 relatively long-lived SCFs. Propagation speeds of the latter were then calculated from phase differences between oscillations detected along multiple cuts with known distances. 
We note that the SCFs under study were detected at different times over the 1~hr time series of images.
Table~\ref{table:stat} summarizes these properties, whose distributions are illustrated in Figure~\ref{fig:statistics}. These include the periods, transverse displacements, and velocity amplitudes of the 134 independent SCFs, as well as the phase speed of the 23 long-lived features.

\begin{table}[!thp]
\caption{Properties of transverse oscillations/waves in slender Ca~{\sc ii}~H fibrils (SCFs) deduced from {\sc Sunrise}/SuFI data.}
\label{table:stat}
\setlength{\tabcolsep}{.5em}                   
\begin{tabular}{l c c c c c}
\hline \hline
Quantity & Range & Mean & Mode & Median & $\sigma$ \tablenotemark{a} \\
\tableline
   Period (s) & 16-199 & 89 & 74 & 83 & 29 \\
   Transverse displacement (km) & 1-91 & 35 & 26 & 33 & 16 \\
   Transverse velocity (km\,s$^{-1}$) & 1.0-4.8 & 2.5 & 2.1 & 2.4 & 0.8 \\
   Phase speed (km\,s$^{-1}$) & 1-70 & 15 & 5 & 9 & 14 \\
\tableline
\vspace{-1mm}
\end{tabular}
\hspace{1mm}
\vspace{0.7mm}
\textbf{Notes.}
\footnotetext[1]{ Standard deviation of the distributions.}
\end{table}

We note that individual disturbances often include a range of properties at their various spatial locations and/or at different times. Hence, in addition to different SCFs, the wide range of each quantity may also include the somewhat different properties of oscillations in individual SCFs (i.e., multiple periods observed in a SCF at different times). Thus, the distributions of period, transverse displacement, and velocity amplitude, shown in Figure~\ref{fig:statistics}(a)--(c), include about 400 individual points. The histogram of the phase speed, although from 23 events, includes about 1400 data points from wavelet cross-power spectra of several pairs of transverse perturbations (the fluctuations were studied at several cuts perpendicular to the axis of each of the 23 SCFs).

Transverse oscillations in the SCFs are found to have a broad range of periods and phase speeds. These also include the propagation of waves in individual SCFs (see, e.g., Figures \ref{fig:xrt} and \ref{fig:xrt2} where the measured oscillations in a SCF have a wide range of periods and phase relationships). This suggests that the observed waves are incoherent (i.e., each fibril oscillates independently).

\subsection{Energy Budget}
\label{subsec:energy}

The energy flux transported by transverse waves can be estimated using

\begin{equation}
	E_{k} = \frac{1}{2}\, f \left ( \rho_{i}+\rho_{e} \right ) \, v^{2}\, c_{ph}\,,
	\label{equ:energy_kink}
\end{equation}
\noindent
if they are MHD kink waves \citep{Van-Doorsselaere2014}. In Equation (\ref{equ:energy_kink}), $\rho_{i}$ and $\rho_{e}$ are mass densities inside and outside the oscillating structure (here the SCFs), respectively, $v$ is the observed transverse velocity amplitude, $f$ is the density filling factor of discrete flux tubes, and $c_{ph}$ is the determined phase speed, respectively. The factor 1/2 reflects the time-averaged energy over one period. Obviously, misidentification of the wave mode (with a non-unity filling factor) results in over/underestimation of energy flux \citep{Van-Doorsselaere2008,Goossens2013}.

Assuming that the SCFs represent flux tubes (which is not certain), the observed transverse oscillations, which are a result of the displacement of the SCFs' axes, can be interpreted as MHD kink waves (see Section~\ref{sec:conclusions} for further discussions).

By combining the observed velocity amplitude and phase speed of 2.4 and 9~km/s, respectively, with a plasma mass density of $\rho_{i}\approx10^{-6}$~kg\,m$^{-3}$ (in the low chromosphere; at height $\approx700$~km from the visible surface) from a realistic radiative MHD simulation~\citep{Carlsson2016}, we roughly estimate the energy flux of the kink waves in the SCFs to be $E_{k}\approx15$~kW/m$^2$. We took $\rho_{i}/\rho_{e}=1$. Depending on whether the magnetic field is larger in the fibrils or outside them, this ratio will be less than or larger than unity. The energy flux for transverse waves will be smaller or larger depending on the ratio of densities between inside and outside the fibrils, but should, for reasonable assumptions of density contrast, not differ by more than a factor of three from the value deduced here. We approximated the filling factor of the SCFs to be $\approx0.3$. This was determined from clean images where solar background, noise, and non-interesting brightenings were removed from the observed data (see Section~\ref{subsec:restoration}). The ratio of the area filled by the SCFs to the total area (averaged over the entire time series of images) was taken as the filling factor. This would, however, be a lower limit for the filling factor since many fibrils were excluded during the several steps of image restoration.

Also, we assume that the (magnetized) gas between the fibrils is not oscillating, which is unlikely, as the shaking fibril should also excite oscillations in that gas. The assumption of a kink wave requires that the density (and/or the magnetic field strength) inside the fibrils is different from that outside the fibrils. In the lower chromosphere, it is unclear to what extent such differences occur, or if the fibrils are mainly visible because of, e.g., a horizontal temperature inhomogeneity in the lower chromosphere leading to an inhomogeneous optical depth scale where the Ca~{\sc ii}~H$_{2}$ peaks are formed.

\vspace{2mm}

\section{Discussion and Conclusions}
\label{sec:conclusions}

We have presented, for the first time to our knowledge, observations of transverse oscillations in SCFs. Seeing-free observations by the SuFI instrument onboard the {\sc Sunrise} 1~m solar telescope in a narrowband Ca~{\sc ii}~H filter have provided us with an excellent long time series of images sampling the lower solar chromosphere. The observed area was part of active region AR11768. The strongly enhanced Ca~{\sc ii}~H brightness in the ROI (see Figure~\ref{fig:obs}) indicates a significant enhancement of the H$_{2V}$/H$_{2R}$ emission in the core of Ca~{\sc ii}~H (due to the enhanced temperatures in the presence of strong magnetic fields; \citealt{Solanki1991,Beck2013}) so that the observed radiation contains significant contributions from the lower chromosphere \citep{Jafarzadeh2017b}.

Our analysis of the {\sc Sunrise}/SuFI Ca~{\sc ii}~H data revealed that the transverse oscillations are ubiquitous in the lower solar chromosphere. To characterize the oscillations reliably, we limited our study to a relatively small number of isolated SCFs not affected by, e.g., the superposition of several SCFs and/or with other brightenings. Most SCFs are visible only for a short time at a stretch, are highly dynamic, and display a high number density in the studied area.

We found a wide range of periods, transverse displacements, velocity amplitudes, and phase speeds (when possible) for the oscillations/waves along the SCFs (see Figure~\ref{fig:statistics}), with median values of 83~s, 33~km, 2.4~km/s, and 9 km/s, respectively (see Table~\ref{table:stat} for all statistical values).

\begin{table*}[!htp]
\caption{Comparison of Properties of Transverse Oscillations/Waves in Our Slender Ca~{\sc ii}~H Fibrils (SCFs) with those in other Chromospheric Fibrillar Structures in the Literature\tablenotemark{a}.}
\label{table:comp}
\setlength{\tabcolsep}{.4em}                   
\begin{tabular}{l c c c c c c c c c c c c c c}
\hline \hline
References & Telescope & Passband & Event\tablenotemark{c} & \multicolumn{2}{c}{Period} && \multicolumn{2}{c}{Transverse} && \multicolumn{2}{c}{Velocity} && \multicolumn{2}{c}{Phase} \\
 &  & (spectral &  & \multicolumn{2}{c}{(s)} && \multicolumn{2}{c}{displacement (km)} && \multicolumn{2}{c}{amplitude (km\,s$^{-1}$)} && \multicolumn{2}{c}{speed (km\,s$^{-1}$)} \\
\cline{5-6} \cline{8-9} \cline{11-12} \cline{14-15}
 &  & resolution [\AA]\tablenotemark{b}) &  & Range & Med. && Range & Med. && Range & Med. && Range & Med.\\
\tableline

This study & {\sc Sunrise} & Ca~{\sc ii}~H (1.1)\tablenotemark{d} & SCF\tablenotemark{e} & 16--199 & 83 && 1--91 & 33 && 1--4.8 & 2.4 && 1--70 & 9 \\

\citet{Morton2014} & DST & H$\alpha$ (0.25)\tablenotemark{d} & Fibril\tablenotemark{e} & 25--620 & 185 && 25--320 & 107 && 0--18 & 5.4 && ... & ... \\ %

\citet{Morton2013} & DST & H$\alpha$ (0.25)\tablenotemark{d} & Fibril\tablenotemark{g} & 10--500 & 157 && 0--400 & 98 && 1--15 & 5.8 && ... & ... \\ 

\citet{Sekse2013} & SST & Ca~{\sc ii}~8542 (0.11)\tablenotemark{d} & RBE\tablenotemark{g} &  &  &&  &  &&  &  && ... & ... \\ %

\citet{Kuridze2012} & DST & H$\alpha$ (0.25)\tablenotemark{d} & Mottle\tablenotemark{g} & 70--280 & 165 && 100--400 & 200 && 3--18 & 8.0 && 40--110 & ... \\ 

\citet{Pereira2012} & \textit{Hinode} & Ca~{\sc ii}~H (3.0)\tablenotemark{f} & Sp.II\tablenotemark{g} & ... & ... && 15--1461 & 189 && 2--55 & 13 && ... & ... \\ %
 &  &  & Sp.II\tablenotemark{e} & ... & ... && 0--2019 & 251 && 0--58 & 11.5 && ... & ... \\ %

\citet{Morton2012} & DST & H$\alpha$ (0.25)\tablenotemark{d} & Fibril\tablenotemark{g} & 180--210 & ... && 100--800 & 340 && 0--21 & 7.2 && 50--90 & ... \\ 

\citet{Okamoto2011} & \textit{Hinode} & Ca~{\sc ii}~H (3.0)\tablenotemark{f} & Sp.II\tablenotemark{g} & 4--200 & 45 && 7--370 & 55 && 1--29 & 7.5 && 33--500 & 270 \\ 

\citet{Rouppe2009} & SST & H$\alpha$ (0.06)\tablenotemark{d} & RBE\tablenotemark{g} & ... & ... && 0--1000 & 320 && 0--21 & 8.1 && ... & ... \\ %

\citet{DePontieu2007} & \textit{Hinode} & Ca~{\sc ii}~H (3.0)\tablenotemark{f} & Sp.II\tablenotemark{g} & 150--350 & ... && 0--1450 & 450 && 0--30 & 11.7 && ... & ... \\ 

\tableline
\vspace{-1mm}
\end{tabular}
\hspace{1mm}
\vspace{0.7mm}
\textbf{Notes.}
\footnotetext[1]{ The range and median of each quantity are given, when available. These values are extracted from , e.g., histograms plotted in the literature, when they are not explicitly provided.}
\footnotetext[2]{ Full Width at Half Maximum (FWHM) of transmission profile of the passband.}
\footnotetext[3]{ Name of the fibrillar structure according to the authors (RBE: Rapid Blueshifted Events; Sp.II: Type~{\sc ii}~spicules)}
\footnotetext[4]{ On-disk observations.}
\footnotetext[5]{ In an active region.}
\footnotetext[6]{ Off-limb observations.}
\footnotetext[7]{ In quiet Sun.}
\end{table*}

The SCFs are found to share some physical characteristics with, e.g., type~{\sc ii}~spicules observed in the upper chromosphere according to~\citet{Gafeira2017a}.
Table~\ref{table:comp} summarizes properties of transverse oscillations (and waves) for various fibrillar structures observed in different chromospheric passbands reported in the literature, along with our results for the SCFs (see \citealt{Jess2015b} for a detailed overview).

In addition to the ranges of different parameters of transverse oscillations in various chromospheric structures, we have also provided the median values of their distributions in Table~\ref{table:comp}, when available. When not explicitly given, they were estimated from plotted distributions. We do not use mean values of the parameters, since they can be susceptible to the influence of outliers. The latter becomes particularly important in the case of fibrillar structures because their precise detection is often a challenge (or they could be mixed with other features of similar appearance). We stress that most of the other studies investigated waves on fibrillar structures located higher in the solar atmosphere (the off-limb studies even much higher), which might explain some of the differences seen in Table 2 and discussed below.  

The SCFs are found to have transverse oscillation with shorter periods than their counterparts in the higher chromospheric layers (with the median period of the former being shorter by a factor of $\approx2$ than the latter). The median period of type~{\sc ii}~spicules reported by \citet{Okamoto2011} is, however, smaller than that of the SCFs (and other chromospheric fibrillar events, including type~{\sc ii}~spicules from the other studies). The authors explained their very short periods as a result of their method of analysis.

Transverse displacement, velocity amplitude and phase speed of the SCFs studied here are significantly smaller than those of the upper chromospheric elongated features (i.e., H$\alpha$ fibrils, type~{\sc ii}~spicules, mottles and H$\alpha$/Ca~{\sc ii}~8542~\AA\ RBEs). Transverse displacements of the relatively short-period type~{\sc ii}~spicules of \citet{Okamoto2011} are, however, comparable to (but slightly larger than) those we found for the SCFs (that is also likely to be an influence of their detection method). One reason for this discrepancy, even with structures observed by \textit{Hinode} in the same spectral line is the fact that many of the other observations in Ca~{\sc ii}~H were off-limb, whereas we are observing near solar disk center. This makes the SuFI observations mainly sensitive to much lower levels in the solar atmosphere, where wave parameters, such as wave amplitudes, are expected to be lower due to the higher gas density, assuming the total energy flux in the waves to be the same at both layers. However, the higher spatial resolution of the SuFI data also allows smaller wave amplitudes to be detected.

The structures listed in Table~\ref{table:comp} have been observed in various spectral lines in some cases off-limb so that they sample different heights in the solar chromosphere \citep{Rutten2007,Leenaarts2009,Leenaarts2012}. In the vigorously dynamic Sun even individual lines may obtain a contribution from a large range of heights \citep{Rutten2007}. Hence, the lines used in the various observations may be affected by different local plasma properties, leading to subtle variations in the determined wave parameters~\citep{Jess2015b}. 

Hence, the chromospheric fibrillar structures sampled by the different observations can be of intrinsically different types \citep{Rutten2007}, with more upright fibrils (potentially) channeling energy to the corona and more slanted ones transferring energy mainly within different parts of the chromosphere. The energies are likely generated in the photosphere as a result of the interactions between plasma and the magnetic field. Note that different characteristics of these lines with respect to opacity causes limb observations to represent much greater atmospheric heights compared to those on the disk. This is particularly the case for observations of type~{\sc ii}~spicules in Ca~{\sc ii}~H off the limb. In that observing geometry, the observed features lie in an upper chromospheric layer compared to on-disk observations of, e.g., SCFs, using a similar, or even narrower bandpass. These events may, however, share some apparent and/or physical characteristics (e.g., both types of underlying fibrils are observed as short-lived, dynamic, thin and bright elongated features in the Ca~{\sc ii}~H passband). \citet{Jafarzadeh2017b} showed that the {\sc Sunrise}/SuFI SCFs map the magnetic fields in the low solar chromosphere.

The observed transverse oscillations in the SCFs can clearly be described as waves propagating along the axes of the SCFs (when phase relationship between different locations along the fibril could be studied). 
In principle, both bulk Alfv\'{e}n and kink waves can lead to transverse displacements.
While some authors have interpreted transverse oscillations in chromospheric (and/or coronal) fibrillar structures, such as type~{\sc ii}~spicules, as Alfv\'{e}n waves \citep[e.g.,][]{DePontieu2007,Tomczyk2007,McIntosh2011}, others have debated such interpretations and have explained them as a result of MHD kink waves \citep[e.g.,][]{Van-Doorsselaere2008,Goossens2013,Van-Doorsselaere2014}. 
In any case, they are Alfv\'{e}nic, i.e., largely incompressible transverse oscillations propagating along the field lines, whose phase speed differs from the Alfv\'en speed only due to the inhomogeneity of the atmosphere (see, e.g., \citealt{Priest2014} for a detailed description of the terminology used here).
The exact wave mode likely to be excited depends on the physical conditions in the SCFs and their surroundings, which are still largely unexplored. The bulk (shear) Alfv\'{e}n waves only exist in a homogeneous medium \citep{Nakariakov2005,Goossens2013}, that is likely not the case in the solar atmosphere.

The SCFs are possibly a manifestation of plasma jets along horizontal flux tubes, thus, according to the MHD wave theory, the kink mode is the most appropriate interpretation of their swaying motions~\citep{Kukhianidze2006,Erdelyi2007b,Zaqarashvili2007}. 
Note, however, that the fact that brightenings along the fibrils start more often inside the bulk  of the fibrils than at one of their edges speaks against the interpretation of the fibrils as plasma jets. The incoherent oscillations among different SCFs disputes the option of  global (bulk) Alfv\'{e}n waves \citep{Zaqarashvili2009} (in addition to the assumption of the non-uniform medium), although it is unclear how large the coherence length must be to qualify as an Alfv\'en wave. A closer inspection would be necessary to confirm whether such incoherency is also consistent between neighboring SCFs.
Note that incompressible, torsional Alfv\'{e}n waves cannot displace flux tubes back and forth, hence they are unlikely to be responsible for the observed oscillations \citep{Ploner1999,Erdelyi2007b,Van-Doorsselaere2008,Jess2009}.

The MHD kink waves can be triggered by granular buffeting of footpoints of magnetic flux tubes at the base of the photosphere \citep[e.g.,][]{Spruit1981a,Choudhuri1993b,Hasan2008} or by small-scale magnetic reconnection \citep{He2009}. The waves could also be generated as a result of wave coupling at plasma-$\beta =1$ regions \citep[e.g.,][]{Bogdan2003,Roberts2004}.

Excursions in the upper chromospheric fibrillar structures (such as mottles, spicules, and H$\alpha$ dynamic fibrils) have been suggested to be produced by chromospheric shock waves \citep{DePontieu2004a,Hansteen2006}. The chromospheric shocks are produced with different strengths and periods that consequently result in a wide range of observed properties of transverse oscillations \citep{Hansteen2006}. This mechanism could also be relevant for the origin of swaying motions in the SCFs, as shocks are also present in the lower chromosphere (e.g. \citealt{Wedemeyer2004}).

We note that the ROI, within which the SCFs were studied, was located outside pores, but in their immediate vicinity (see Figure~\ref{fig:obs}). Therefore, some of the waves in our SCFs, particularly close to the left and bottom edges of the ROI, could be excited within the pores, or could at least be influenced by pore waves \citep[e.g.,][]{Morton2012b,Khomenko2015}.

The observed high-frequency kink waves were estimated to carry an energy flux of order 15~kW\,m$^{-2}$, which is comparable with what is needed to heat the (lower) solar chromosphere \citep{Withbroe1977,Vernazza1981}. Our observations cannot, however, clarify whether this energy is dissipated in the sampled atmospheric layers. The apparently almost horizontal orientation of the majority of the SCFs studied here, many of which connect back to the solar surface as low-lying loops (see \citealt{Jafarzadeh2017b}), suggest that the waves propagating along them may not reach the upper atmosphere. Rather, they play a role in heating their local surroundings, if they release their energy at these layers. \citet{Jafarzadeh2017b} showed with the help of the magnetostatic solution of \citet{Wiegelmann2017} that the shorter SCFs map fields that are more vertical. A number of the transverse waves we have identified are found along these short fibrils. These waves are likely capable of traveling upwards into the upper atmosphere and may even reach to the corona, contributing to the heating there.

To determine which mechanism can precisely explain the nature of bright and highly dynamic SCFs, analysis of high-resolution observations in a narrower Ca~{\sc ii}~H passband (e.g., from the upcoming SST/CHROMIS instrument) as well as realistic MHD simulations of this line will be necessary.

\acknowledgements
The German contribution to {\sc Sunrise} and its reflight was funded by the Max Planck Foundation, the Strategic Innovations Fund of the President of the Max Planck Society (MPG), DLR, and private donations by supporting members of the Max Planck Society, which is gratefully acknowledged. The Spanish contribution was funded by the Ministerio de Econom\'{\i}a y Competitividad under Projects ESP2013-47349-C6 and ESP2014-56169-C6, partially using European FEDER funds. The HAO contribution was partly funded through NASA grant number NNX13AE95G. This work was partly supported by the BK21 plus program through the National Research Foundation (NRF) funded by the Ministry of Education of Korea. S.J. receives support from the Research Council of Norway. S.J. is grateful to E. Priest for useful discussions.

\bibliographystyle{aa}
\bibliography{Shahin}

\end{document}